\documentclass[aps, prl, twocolumn, floatfix, superscriptaddress, reprint, 10pt]{revtex4-1} 
\usepackage{mhchem}

\newcommand{\br}{\mathbf{r}}
\newcommand{\bk}{\mathbf{k}}
\newcommand{\bc}{\mathbf{c}}
\newcommand{\bK}{\mathbf{K}}
\newcommand{\bS}{\mathbf{S}}

\newcommand{\bw}{\mathbf{w}}

\newcommand{\bx}{\mathbf{x}}

\newcommand{\CX}{\mathcal{X}}

\newcommand{\loss}{\ell}

\usepackage{bm}
\usepackage{graphicx}
\usepackage[usenames]{color}
\usepackage{microtype}
\usepackage{amsmath}
\usepackage{amssymb}
\usepackage{xspace}
\usepackage{footnote}
\usepackage{relsize}
\usepackage{hhline}
\usepackage{verbatim}
\usepackage[normalem]{ulem}

\newcommand{\be}{\begin{equation}}
\newcommand{\ee}{\end{equation}}
\definecolor{mygray}{gray}{0.5}

\usepackage{floatrow}
\newfloatcommand{capbtabbox}{table}[][\FBwidth]

\begin{document}

\title{A Transferable Machine-Learning Model of the Electron Density
}

\author{Andrea Grisafi}
\affiliation{Laboratory of Computational Science and Modeling, IMX, \'Ecole Polytechnique F\'ed\'erale de Lausanne, 1015 Lausanne, Switzerland}

\author{David M. Wilkins}
\affiliation{Laboratory of Computational Science and Modeling, IMX, \'Ecole Polytechnique F\'ed\'erale de Lausanne, 1015 Lausanne, Switzerland}

\author{Benjamin A. R. Meyer}
\affiliation{Laboratory for Computational Molecular Design, Institute of Chemical Sciences and Engineering,\'Ecole Polytechnique F\'ed\'erale de Lausanne, CH-1015 Lausanne, Switzerland}

\author{Alberto Fabrizio}
\affiliation{Laboratory for Computational Molecular Design, Institute of Chemical Sciences and Engineering,\'Ecole Polytechnique F\'ed\'erale de Lausanne, CH-1015 Lausanne, Switzerland}

\author{Clemence Corminboeuf}
\affiliation{Laboratory for Computational Molecular Design, Institute of Chemical Sciences and Engineering,\'Ecole Polytechnique F\'ed\'erale de Lausanne, CH-1015 Lausanne, Switzerland}

\author{Michele Ceriotti}
\affiliation{Laboratory of Computational Science and Modeling, IMX, \'Ecole Polytechnique F\'ed\'erale de Lausanne, 1015 Lausanne, Switzerland}

\begin{abstract}
The electronic charge density plays a central role in determining the behavior of matter at the atomic scale, but its computational evaluation requires demanding electronic-structure calculations. 
We introduce an atom-centered, symmetry-adapted framework to machine-learn the valence charge density based on a small number of reference calculations. The model is highly transferable, meaning it can be trained on electronic-structure data of small molecules and used to predict the charge density of larger compounds with low, linear-scaling cost. 
Applications are shown for various hydrocarbon molecules of increasing complexity and flexibility, and demonstrate the accuracy of the model when predicting the density on octane and octatetraene after training exclusively on butane and butadiene. 
This transferable, data-driven model can be used to interpret experiments, initialize electronic structure calculations, and compute electrostatic interactions in molecules and condensed-phase systems. \end{abstract}

\maketitle

\section{Introduction}

The electron density $\rho(\br)$ is a fundamental property of atoms, molecules and condensed phases of matter. 
$\rho(\br)$ can be measured directly by high-resolution electron diffraction~\cite{koritsanszky_chemical_2001, gatti_modern_2012} and transmission electron microscopy~\cite{Meyer2011}, and can be analyzed to identify covalent and non-covalent patterns.~\cite{coppens_charge_2005,becke_1990,johnson_revealing_2010,de_silva_simultaneous_2014, Pastorczak2017} Based on density-functional theory (DFT), in the framework of the first Hohenberg-Kohn theorem\cite{parr1994}, knowledge of $\rho(\br)$ gives access, in principle, to any ground-state property.
Especially for large systems, however, the computation of $\rho(\br)$ requires considerable effort, involving the solution of an electronic structure problem with a more or less approximate level of theory.
Sidestepping these calculations and directly accessing the ground-state electron density for a given configuration of atoms would have broad implications,
including
a more accurate determination of atomic positions within X-ray atomic structure refinements, real-time visualization of chemical fingerprints based on the electron density, acceleration of DFT calculations by providing a better initial guess for the self-consistent calculation, and an exact treatment of the electrostatic interactions within an atomistic simulation.  

Recently, a landmark paper by Brockherde \emph{et al.} showed that it is possible to predict the ground-state electron density in a way that mimics the Hohenberg-Kohn mapping between the nuclear potential and the density~\cite{burke2017}. A smoothed representation of the nuclear potential was used as a fingerprint to describe molecular configurations, and to carry out individual predictions of the expansion coefficients of $\rho(\br)$ represented in a plane-wave basis. 
Though in principle very effective, the structure of the model imposes significant constraints on its transferability to large and flexible systems. Indeed, the use of a global representation of the structure, and of an orthogonal basis to expand the density, means that the model is limited to interpolation between conformers of relatively rigid, small molecules.

In this paper, we show how to overcome these limitations by constructing a machine-learning model of the valence electron density that can be used on both large and flexible systems, and that is transferable enough to predict the density on large molecules based on training on smaller compounds. This is possible, in a nutshell, thanks to the combination of a local basis set to represent $\rho(\br)$ and a recently introduced regression model which allows us to interpolate the local components of $\rho(\br)$ in a symmetry-adapted fashion. 

The method is tested on the carbon series C$_2$, C$_4$ and C$_8$ of both fully saturated and unsaturated hydrocarbons, having increasing complexity because of the exponentially growing number of conformers. In particular, interpolation of the electron density is first shown for ethene (C$_2$H$_4$), ethane (C$_2$H$_6$), butadiene (C$_4$H$_6$) and butane (C$_4$H$_{10}$). As a major result, the electron density of the corresponding C$_8$ molecules, namely octa-tetraene (C$_8$H$_{10}$) and octane (C$_8$H$_{18}$), is instead predicted by extrapolating the information learned on the local environments of the corresponding C$_4$ molecules. 

\section{Theory and Methods}

\subsection{Symmetry-Adapted Gaussian Process Regression for the Charge Density}

Several widely-adopted machine-learning schemes applied to materials rely on an additive decomposition of the target property in atom-centered contributions~\cite{behl-parr07prl,bart+10prl,smit+17cs}. These approaches are very effective in achieving transferability across systems of different composition and size.
An additive ansatz is justified by the exponential decay of the electronic density matrix (the so-called nearsightedness principle~\cite{prod-kohn05pnas}), which also underlies a plethora of linear-scaling, embedding and fragment decomposition electronic structure methods~\cite{yang91prl,gall-parr92prl,goed99rmp,ceri+08jcp,marz-vand97prb,fedorov_extending_2007,merz_using_2014,walker_molecular_1993,meyer_libraries_2016}.
Many methods exist to decompose the density in atom-centered contributions~\cite{hirshfeld_bonded-atom_1977,Bader1990}, which however share a degree of arbitrariness. \cite{gonthier_quantification_2012}
Given that only the total density $\rho(\br)$ is physically meaningful, using a preliminary decomposition in atomic contributions, and then using those as machine-learning targets, would impose a number of unnecessary (and largely arbitrary) constraints. For this reason, we introduce locality only by expanding the density as a sum of atom-centered basis functions,
\begin{equation}\label{decomposition}
\rho(\br) = \sum_i \rho_i(\br) = \sum_{ik} c^i_{k}\, \phi^i_{k}(\br)= \sum_{ik} c^i_{k}\, \phi_{k}(\br - \br_{i}),
\end{equation}
where $k$ runs over the basis functions centered on each atom, and atoms of different species can have different kinds of functions. The combination coefficients $c^i_{k}$ are the target property that we aim to machine-learn in order to provide a prediction of the electron density based exclusively on the knowledge of the positions of the nuclei.
From an atom-centred description, it is natural to factorize each basis function $\phi_{k}(\br - \br_{i})$ into a product of radial functions~$R_{n}(r_i)$ and spherical harmonics~$Y_{m}^{l}(\hat{\br}_{i})$ (with $r_i=\left|\br-\br_i\right|$ and $\hat{\br}_i=(\br-\br_i)/r_i$). 
The subscript $k$ refers to the combination $nlm$, and we will use the compact or the extended notation based on convenience. 
For every atom-centered environment $\CX_i$, which defines the structure of a neighborhood of atom $i$, and for each radial function $R_{n}$, the coefficients can be grouped according to their value of angular momentum $l$ in a set of spherical multipoles $\bc^i_{nl}$ of dimension $2l+1$, which transform as vector spherical harmonics $\mathbf{Y}_{l}$ under a rigid rotation of the environment. 
This choice has the advantage of highlighting the tensorial nature of the density components, meaning that a great portion of the variability of $\bc^i_{nl}$ can be attributed to the orientation of the local environments $\mathcal X_i$, rather than to an actual structural distortion of the molecule.

Dealing with the regression of tensorial properties raises non-trivial issues in terms of setting up an effective machine-learning model that takes into account the proper covariances in three dimensions.
For rigid molecules, one could eliminate this geometric variability by expressing the coefficients in a fixed molecular reference frame, analogously to what has already been done in the context of electric multipoles and response functions~\cite{Bereau2015,liang2017}.
However, this strategy cannot be
extended to arbitrarily complex and flexible molecules, where a single body-centred reference frame is in general ill-defined. 
As shown recently, Gaussian process regression can be modified to naturally endow the machine-learning model of vectors~\cite{glielmo} and tensors of arbitrary order~\cite{grisafi2018} with the symmetries of the 3D rotation group SO(3). Within this method, called symmetry-adapted-Gaussian-process-regression (SA-GPR), the machine-learning prediction of the tensorial density components is:
\begin{equation}
c^i_{nlm}(\bx)= \sum_{j\in M} \sum_{|m'|<l} \ k^l_{mm'}(\mathcal{X}_{i},\mathcal{X}_j)\ x^j_{nlm'} \delta_{\alpha_i\alpha_j}
\label{eq:krr-model}
\end{equation}
In this expression, $\bk^l(\mathcal X_i,\mathcal X_j)$ is a rank-2 kernel matrix of dimension $(2l+1)\times(2l+1)$ that expresses, at the same time, both the structural similarity and the geometric relationship between the atom-centered environment $\mathcal X_i$ of the target molecule and a set $M$ of reference environments~$\mathcal X_j$. The (tensorial) regression weights $x^j_{nlm'}$ are determined from a set of $N$ training configurations and their associated electron densities. 

According to Eq.~\eqref{eq:krr-model}, the prediction of the density expansion coefficients $c^i_{nlm}(\bx)$ is performed independently for each radial channel $n$, angular momentum value $l$ and atomic species $\alpha$. 
However, working with a non-orthogonal basis implies that the density components belonging to different atoms of the molecule are not independent of each other. 
One can indeed evaluate the projections of the density on the basis functions
\begin{equation}
w^i_{k}=\left<\rho | \phi^i_{k}\right>=\int \mathrm{d}\br\ \rho(\br)\ \phi_{k}(\br -\br_i),
\end{equation}
but these differ from the expansion coefficients $c^i_k$. 
In fact, $\mathbf{w}$ and $\mathbf{c}$ are related by $\mathbf{S}\mathbf{c}=\mathbf{w}$, where $S^{ii'}_{kk'}=\left<\phi^i_{k}| \phi^i_{k'}\right>$ is the overlap between basis functions.
For a given density, the coefficients could therefore be determined by inverting $\mathbf{S}$, so that each individual $nl\alpha$ component could be machine-learned separately.
We observed, however, that doing so led to poor regression performance and unstable predictions. Applying $\bS^{-1}$ on $\bw$ corresponds to a partitioning of the charge which is, most of the times, affected by numerical noise.
This is connected to the fact that $\mathbf{S}$ is often ill-conditioned, and so 
small numerical errors in the determination of $\bw$ translate into large instabilities in the coefficients $\bc$, making it hard for the machine-learning algorithm to find a unique relationship between the density components and the nuclear coordinates of the molecule. 
To avoid this issue and improve the accuracy of the physically-relevant total density, the basis set decomposition and the construction of the machine-learning model need to be combined into a single step. This essentially consists in building a regression model that, of the many nearly equivalent decompositions of $\rho$, is able to determine the one which best fits the target density associated with a given structure.

The problem can be cast into a single least-square optimization of a loss function that measures the discrepancy between the reference and the model densities,
\begin{equation}\label{eq:loss}
    \loss({\bx}) = \sum_{\mathcal A \in N} \int d\br\ \left|\rho_{\mathcal A}(\br) - \sum_{i\in \mathcal A}\sum_k c^i_{k}(\bx)\, \phi_{k}(\br - \br_{i}) \right|^2 + \eta \left|\bx\right|^2.
\end{equation}
Here the index $N$ runs over the training set while $i$ runs over the environments of a given training structure. The second term in the loss is a regularization, which avoids overfitting. In this context, $\eta$ represents an adjustable parameter that is related to the intrinsic noise of the training dataset.
The coefficients $\bc$ depend parametrically on the regression weights $\bx$ via Eq.~\eqref{eq:krr-model}; by differentiating the loss with respect to $x^j_{nlm}$ one obtains a set of linear equations that make it possible to evaluate the weights in practice. In compact notation, the solution of this problem reads
\begin{equation}\label{eq:regression_formula}
\bx = \left(\bK^T \bS \bK + \eta \mathbf{ 1}\right)^{-1} \bK^T \bw
\end{equation}
where $\bx$ and $\bw$ are vectors containing the regression weights and the density projections on the basis functions, while $\bK$ and $\bS$ are sparse matrix representations containing the symmetry-adapted tensorial kernels and the spatial overlaps between the basis functions. The details of this derivation and the resulting expressions are given in the SI. 
It should, however, be stressed that the final regression problem is highly non-trivial. The kernels that involve environments within the same training configuration are coupled by the overlap matrix, so that all the regression weights $\bx$ for different elements, radial and angular momentum values must be determined simultaneously.
An efficient implementation of a ML model based on Eqn.~\eqref{eq:regression_formula} requires the optimization of a basis set for the expansion, the evaluation of $\rho(\br)$ on dense atom-centered grids, the sparsification of the descriptors that are used to evaluate the kernels, and the determination of a diverse, minimal set of reference environments $\CX_j$. 
All of these technical aspects are discussed extensively in the~SI. 

\subsection{Benchmark Dataset}

As a demonstration of this framework we consider hydrocarbons, using a dataset of 1000 independent structures of ethene, ethane, butadiene and butane. Atomic configurations are generated by running replica exchange molecular dynamics (REMD) simulations at the density functional tight binding level~\cite{Elstner2014}, using a combination of the DFTB+~\cite{Aradi2007} and i-PI~\cite{ceri+14cpc} simulation software.~\cite{petr+15jcc} A diverse set of 1000 configurations was then extracted from the replica at $T=300$~K by farthest point sampling (FPS), based on the SOAP metric~\cite{de+16pccp,imba+18jcp}.
For each selected configuration we computed the electron density at the DFT/PBE level with SBKJC effective core potentials. Further details of the dataset construction are given in the SI.

The problem of representing a charge density in terms of a non-orthogonal localized basis set shares many similarities with that of expanding the wavefunction. 
For this reason, we resort to many of the tricks used %
in quantum chemistry codes, including the use of Gaussian type orbitals (GTOs) to compute the basis set overlap analytically, and the contraction of 12 regularly spaced radial GTOs down to 4 optimized functions.
We find that angular momentum channels up to \emph{f} functions are needed to obtain a decomposition error around 1\%{} for the density.
The coefficients of the contraction are optimized to minimize the mean charge decomposition error and the condition number of the overlap matrix for the four molecules~\cite{Joost2007}, as discussed in the SI. 
A systematic analysis of the interplay between the details of the basis set and the performance of the ML model goes beyond the scope of this work. It is likely however that substantial improvements of this approach could be achieved by further optimization of the basis. 

\begin{figure}[htb!]
 \includegraphics[width=1.\textwidth]{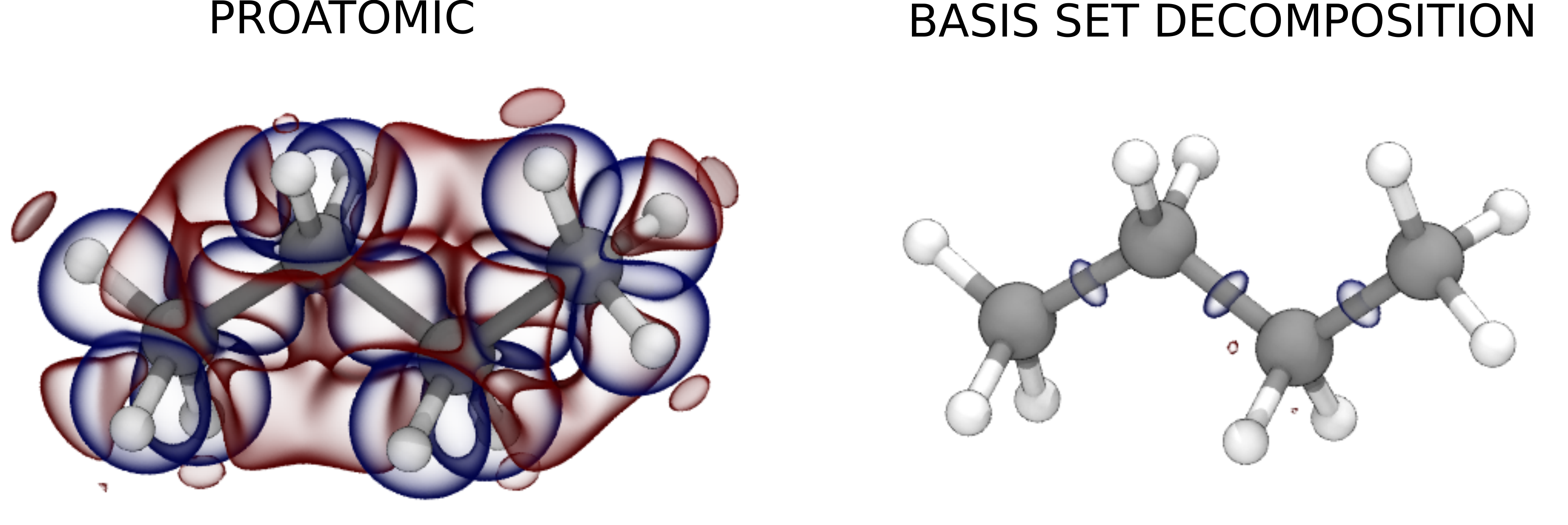}\\[1cm]
\begin{tabular}{c|cccc}
\hline\hline
$\left<\varepsilon_\rho\right>$(\%) & \ce{C2H4} & \ce{C2H6} &\ce{C4H6} & \ce{C4H10} \\
\hline
Proatomic & 18.06 & 19.23 & 16.79 & 18.13 \\
Basis Set & 1.04 & 1.14 & 0.98 & 1.19\\
\hline\hline
\end{tabular}
\caption{Density errors at different level of representation: (\emph{left}) superposition of isolated atomic densities, (\emph{right}) optimized basis set. Red and blue isosurfaces refer to an error of $\pm$0.005 Bohr$^{-3}$ respectively.
The density errors for the structure depicted are reported in the two panels, while the table reports the mean errors over the whole training set for the \ce{C2} and \ce{C4} molecules.
\label{fig:comparison_basis}}
\end{figure}

\begin{figure}[btp]
 \includegraphics[width=1.\textwidth]{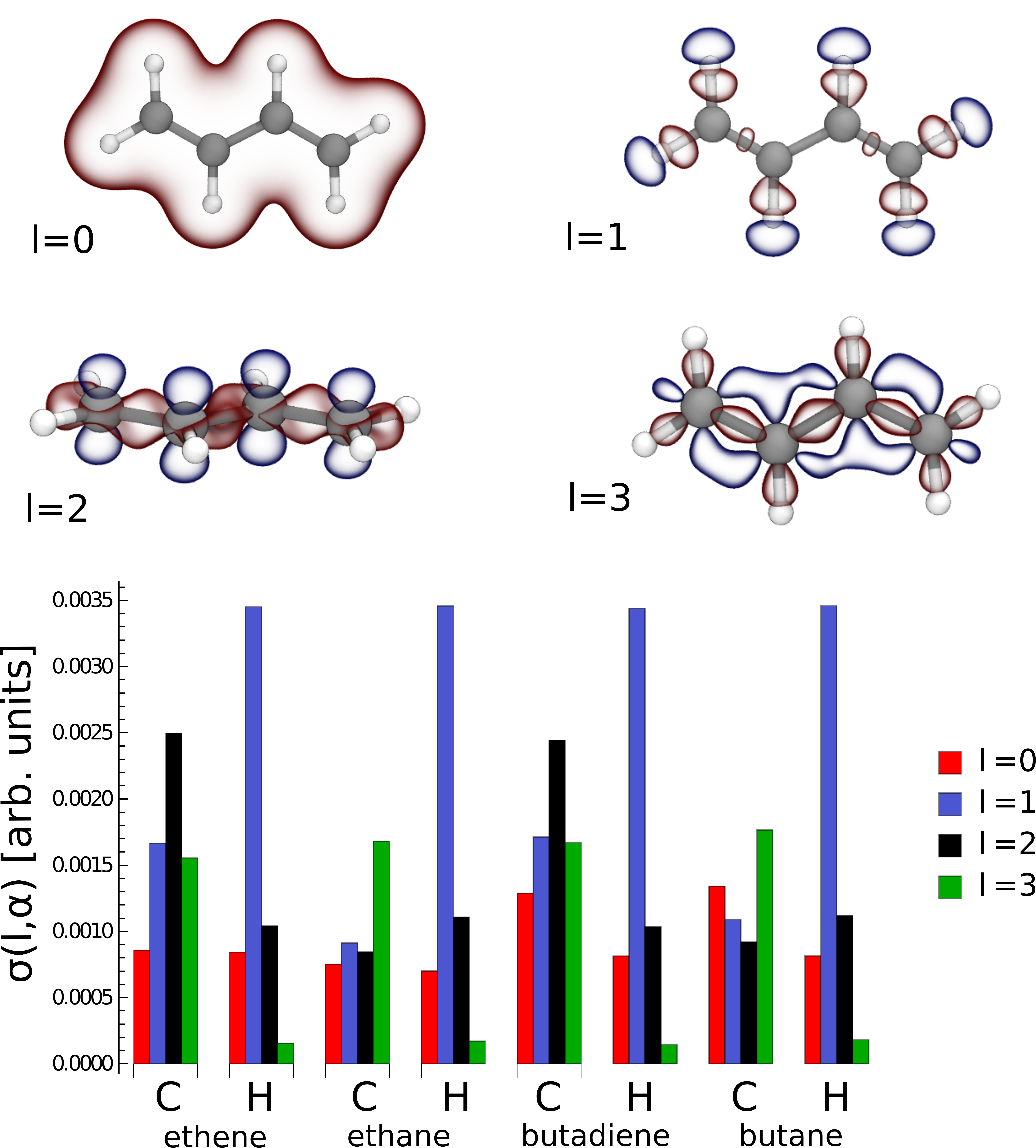}
\caption{(\emph{top}) representation of the angular momentum decomposition of the electron density. Red and blue isosurfaces refer to $\pm 0.01$ Bohr$^{-3}$ respectively. (\emph{bottom}) angular momentum spectrum of the valence electron density of C$_2$ and C$_4$ datasets. The isotropic contributions $l=0$ express the collective variations with respect to the dataset's mean value, while the mean is statistically zero for $l>0$. 
\label{fig:l-spectrum}}
\end{figure}

\section{Results and Discussion}

\subsection{Charge decomposition analysis}

It is instructive to inspect the decomposition of the charge density in terms of the optimized basis, obtained from density projections on the basis functions $\bw$ and the overlap matrix $\bS$ as $\bc = \bS^{-1}\bw$, which corresponds to the best accuracy that can be obtained with a given basis.
With a basis set of 4 contracted radial functions, and angular momentum components up to $l=3$, the typical error in the density decomposition can be brought down to about 1\%{}. In Figure~\ref{fig:comparison_basis} we compare, for the case of a butane molecule, the residual in the expansion with the typical error that can be expected by taking a superimposition of free-atom densities, between 16 and 20\%.

\begin{figure*}[htbp]
\includegraphics[width=1.\textwidth]{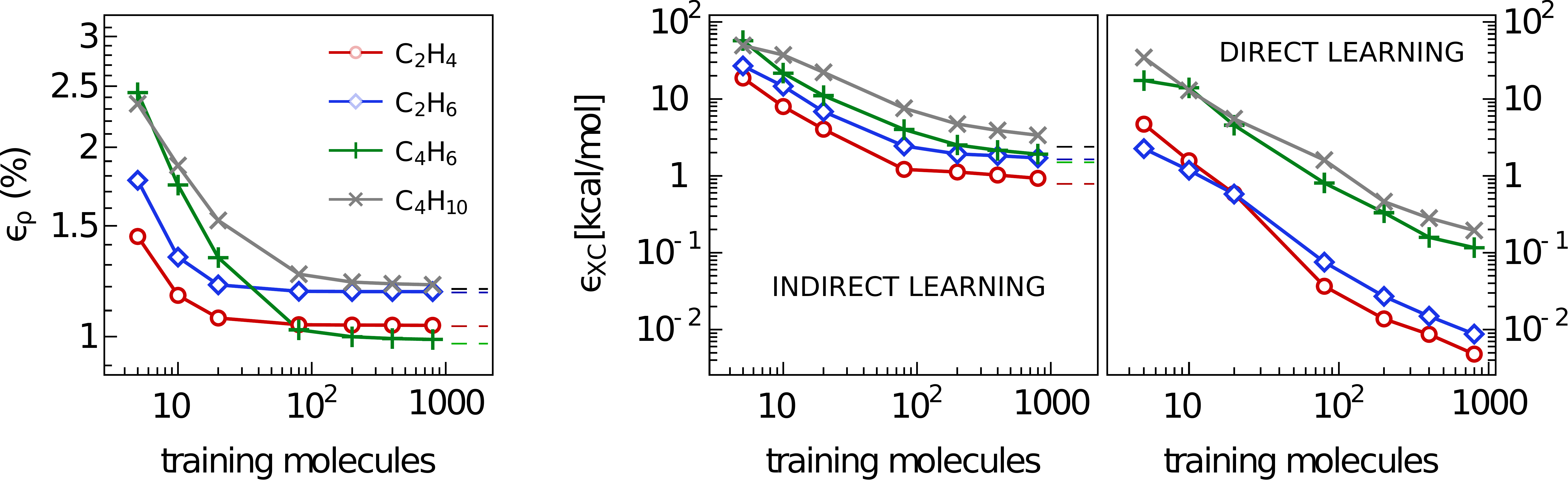}
\caption{Learning curves for C$_2$ and C$_4$ molecules. (\emph{left}) \% mean absolute error of the predicted SA-GPR densities as a function of the number of training molecules. The error normalization is provided by the total number of valence electrons. (\emph{right}) root mean square errors of the exchange-correlation energies indirectly predicted from the SA-GPR densities and directly predicted via a scalar SOAP kernel, as a function of the number of training molecules. Dashed lines refer to the error carried by the basis set representation.
\label{fig:l-curves-carbon}}
\end{figure*}

It is also possible to compute separately the contributions to the charge carried by each angular momentum channel $l$, e.g., $\rho_l(\br)=\sum_{inm} c^i_{nlm}\phi_{nlm}(\br-\br_i)$. 
As exemplified in Fig.~\ref{fig:l-spectrum}, while the isotropic $s$ functions determine the general shape of the density, the $p$ functions primarily describe the gradient of electronegativity in the region close to C--H bonds. Furthermore, the $d$ functions describe the charge modulation associated with the C--C bonds along the main chain as well as the $\pi$-cloud along the conjugated backbone, while the $f$ functions act as a further modulation that captures the non-trivial anisotropy. The figure also shows the collective contribution to the charge variability carried by each angular momentum channel $l$ and atomic type $\alpha$, i.e., $\sigma(l,\alpha)=\sqrt{\sum_{n}\langle|\bc^i_{ln}-\langle \bc^i_{ln} \rangle|^2\rangle_{\alpha_i=\alpha}}$, with the average $\langle\cdot\rangle$ involving all the atoms of the same type included in the dataset. 

Interestingly, after having subtracted the mean atomic density of pure $s$ character, the $l=1$ components largely dominate the charge density variability associated with hydrogen atoms. Functions with $d$ symmetry also carry a substantial contribution, particularly for the carbon atoms of alkenes, while $f$ functions appear to be dominant for carbon atoms of alkanes and almost irrelevant for hydrogen atoms in all the four molecules. 
In comparison to an atom-centered expansion of the wave function $\psi$, the choice of using a larger basis set is justified by the greater complexity in describing an electron density field rather than the $N_\text{e}$/2 occupied orbitals being the solution of an effective single particle Hamiltonian. The need for high angular momentum components can be also justified by the fact that -- even neglecting the overlap between adjacent atoms -- the squaring of $\psi$ that yields $\rho(\br)$ would introduce non-zero components with up to twice the maximum $l$ used to expand the wave-function.

\subsection{Density learning with SA-GPR}

Having optimized the basis set and analyzed the variability of the electron density when expanded in this optimized basis, we now proceed to test the SA-GPR regression scheme. The difficulty of the learning exercise largely depends on the structural flexibility of the molecules. 
Small, rigid systems such as ethene and ethane require little training, and could be equivalently learned through a machine-learning framework based on a pairwise comparison of aligned molecules. 
Butadiene data, containing both \emph{cis} and \emph{trans} conformers, as well as distorted configurations approaching the isomerization transition-state, poses a more significant challenge, due to an extended conjugated system that makes the electronic structure very sensitive to small molecular deformations.
The case of butane is also particularly challenging because of the broad spectrum of intramolecular non-covalent interactions spanned by the many different conformers contained in the dataset. Being fully flexible, this kind of system is expected to benefit most from a ML scheme that can adapt its kernel similarity measure to different orientations of molecular sub-units.
Fig.~\ref{fig:l-curves-carbon} shows the performance of the method in terms of prediction accuracy of the electron density as a function of the number of training molecules. 
The number $M$ of reference environments has been fixed to the 1500 most diverse, FPS-selected, environments contained in each dataset. The convergence with respect to $M$ is discussed in the SI.
The symmetry adapted similarity measure which enters in the regression formula of Eq.~\eqref{eq:regression_formula} is given by the tensorial $\lambda$-SOAP kernels of Ref.~\cite{grisafi2018}. 
This generalizes the scalar ($\lambda=0$) smooth overlap of atomic positions framework~\cite{bartok2013} that has been used successfully for constructing interatomic potentials~\cite{bart+10prl,deringer2017} and predicting molecular properties~\cite{de+16pccp,bart+17sa}. In constructing these kernel functions, we chose a radial cutoff of 4.5~{\AA} for the definition of atomic environments (further details are in the SI). 
Learning curves are then obtained by varying the number of training molecules up to 800 randomly selected configurations out of the total of 1000. The remaining 200 molecules for each of these random selections are used to estimate the error in the density prediction. 

We express the error in terms of the mean
absolute difference between the predicted and  quantum mechanical densities, i.e., $\varepsilon_\rho(\%) = 100 \times \langle\int d\br\ \left|\rho_\text{QM}(\br)-\rho_\text{ML}(\br)\right|\rangle/N_\text{e}$.
The prediction errors of ethene and ethane saturate to the limit set by the basis set representation, which is around 1\%{} for all molecules, with as few as 10 training points.  As expected, given the greater flexibility, learning the charge density of butadiene and butane is more challenging, requiring the inclusion of more than 100 training structures in order to approach the basis set limit.
This level of accuracy (an error which is almost 20 times smaller than that obtained with a superposition of rigid atomic densities, as discussed above) would be sufficient for most applications that rely on the accuracy of the density representation, such as the modelling of X-ray and transmission electron microscopy,\cite{koritsanszky_chemical_2001, gatti_modern_2012,Meyer2011} or the evaluation of density-based fingerprints of chemical interactions~\cite{coppens_charge_2005,becke_1990,johnson_revealing_2010,de_silva_simultaneous_2014, Pastorczak2017}.

Using the predicted $\rho(\br)$ as the basis for a density-functional calculation is more challenging. As a benchmark for this application, we use the SA-GPR predictions for $\rho(\br)$ to evaluate the PBE exchange-correlation functional $E_{XC}[\rho]$ used for the reference quantum-mechanical calculations. 
Depending on the gradient of the density, this quantity is very sensitive to small density variations, especially localized around the atomic nuclei. Fig.~\ref{fig:l-curves-carbon} shows the root mean square error for the exchange-correlation energies~$\varepsilon_\text{XC}$. Using the full set of 800 training molecules, we reach a RMSE of  0.9 and 1.7 kcal/mol for ethene and ethane, 1.9 kcal/mol for butadiene and 3.5 kcal/mol for butane, basically matching the basis set limit. 
It is clear that the ML scheme has the potential to reach higher accuracy with a small number of reference configurations, but a significant reduction of the basis set error is necessary to reach chemical accuracy (roughly 1 kcal/mol RMSE) in the prediction of~$E_\text{XC}$.
At the same time, it is not obvious that computing~$E_\text{XC}$ indirectly, by first predicting the electron charge density, is the most effective strategy to obtain a ML model of DFT energetics. 
As shown in the figure, applying a direct, scalar regression based on conventional SOAP kernels to learn the relationship between the molecular structure and~$E_\text{XC}$ leads to vastly superior performance while requiring a much simpler machine-learning model.

\subsection{Size-extensive extrapolation}

While incremental improvements of the underlying density representation framework are desirable to use the predicted density as the basis of DFT calculations, we can already demonstrate the potential of our SA-GPR scheme in terms of transferability of the model. 
From the prediction formula of Eq.~\eqref{eq:krr-model}, it is clear that no assumption is made about the identity of the molecule for which the electron density is predicted. 
Practically speaking, the regression weights $x^j_{nlm}$ are associated with representative environments that could be taken from any kind of compound, not necessarily the same as that for which the density is being predicted.
As long as the training set is capable of describing different chemical environments, and contains local configurations similar to the ones of our prediction target, accurate densities can be obtained simply by computing the kernels $\mathbf{k}^l(\mathcal{X}_i,\mathcal{X}_j)$ between the environments $\mathcal{X}_i$ of an arbitrarily large molecule and the reference environments $\mathcal{X}_j$. 
The cost of this prediction is proportional to the number of environments, making this method of evaluating the electron charge density strictly linear scaling in the size of the target molecule.

\begin{figure}[htbp]
 \includegraphics[width=1\textwidth]{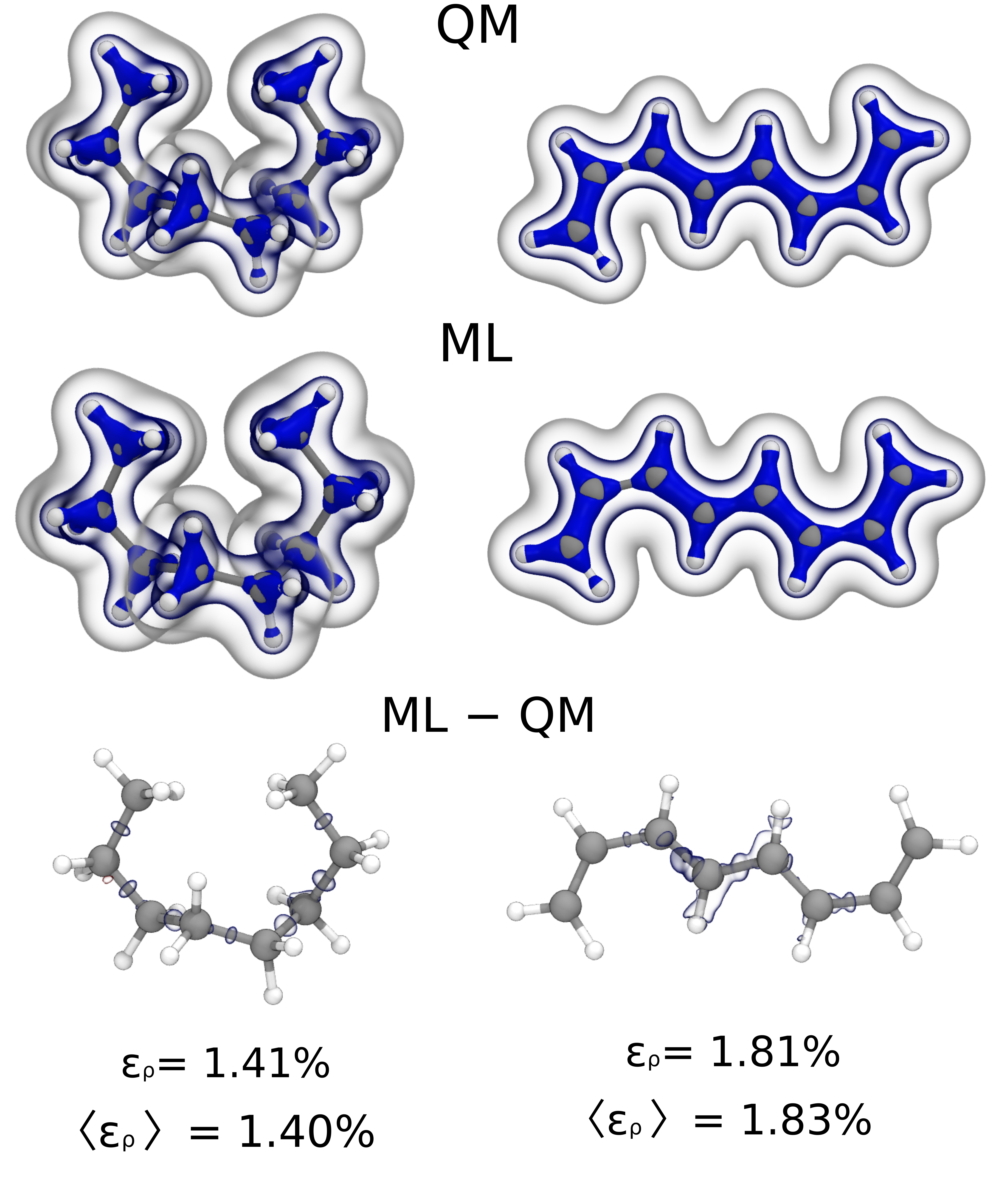}
\caption{Extrapolation results for the valence electron density of one octane (\emph{left}) and one octatetraene (\emph{right}) conformer. (\emph{top}) DFT/PBE density isosurface at 0.25, 0.1, 0.01 Bohr$^{-3}$, (\emph{middle}) machine-learning prediction isosurface at 0.25, 0.1, 0.01 Bohr$^{-3}$, (\emph{bottom}) machine-learning error, red and blue isosurfaces refer to $\pm$ 0.005 Bohr$^{-3}$ respectively. Relative mean absolute errors averaged over 100 conformers are also reported for both cases.
\label{fig:extrapolation}}
\end{figure}

As a proof of concept of this extrapolation procedure, we use environments and training information from the butadiene and butane configurations already discussed to construct the electron density of octatetraene (C$_8$H$_{10}$) and octane (C$_8$H$_{18}$), respectively.
It is important to stress that the transferability is due to the fact that on a local scale the larger molecules are similar to those used for training, and so the prediction is effectively an \emph{interpolation} in the space of local environments. 
This is emphasized by the observation that the optimal extrapolation accuracy is obtained using a machine learning cutoff of $r_\text{cut}=3$~{\AA}, versus a value of $r_\text{cut}=4.5$~{\AA} that was optimal for same-molecule predictions. On a scale larger than $3$~\AA{}, the environments present in C8 molecules differ substantially from those in the corresponding C4 compound, which negatively affects the transferability of the model. 
Ideally, as the training dataset is extended to include larger and larger molecules, this locality constraint can be relaxed until no substantial difference can be appreciated between the prediction accuracy of the interpolated and extrapolated density.

For both octane and octatetraene, the extrapolation is carried out on a challenging dataset made of the 100 most diverse structures extracted by farthest point sampling from the 300 K replica of a long REMD run. When learning on the full dataset of butadiene and butane, we obtain a low density mean absolute error of 1.8\% for octatetraene and of 1.4\% for octane. 
As shown in Fig.~\ref{fig:extrapolation} for two representative configurations, the size-extensive SA-GPR prediction accurately reproduces the structure of the electron density for both octane and octatetraene. Because of the high sensitivity of the electronic $\pi$-cloud to the molecular identity and configuration, major difficulties arise in predicting the electron density of octatetraene, particularly in the middle regions, for which no analogous examples are contained in the butadiene training dataset.

\section{Conclusions}

Machine-learning the electronic charge density of molecular systems as a function of nuclear coordinates poses great technical and conceptual challenges. Transferability across molecules of different size and stoichiometry calls for a scheme based on a local decomposition, which should be performed without relying on arbitrary charge partitioning or discarding the fundamental physical symmetries of the problem. 
The framework we present here overcomes these hurdles by decomposing the density in optimized atom-centered basis functions, exploiting a symmetry-adapted regression scheme to incorporate geometric covariances, and by designing a loss function that relies only on the total charge density as a physically-meaningful constraint. The atom-centered decomposition means the ML model can predict the density of large molecules or condensed phases with a cost that scales linearly with the number of atoms.

We have demonstrated the viability and accuracy of this scheme by learning the ground-state valence electron density of saturated and insaturated hydrocarbons with 2 and 4 carbon atoms, achieving in all cases an error of the order of 1\%{} on the reconstructed density, that is sufficient to enable an order-of-magnitude improvement of the accuracy of methods that usually rely on the superimposition of atomic densities, e.g. in the analysis of X-ray and transmission electron microscopy experiments. 
What is more, models trained on C4 compounds can be used to predict the electronic charge of their larger, C8 counterparts, providing a first  example of the transferability that is afforded by a symmetry-adapted local decomposition scheme.

Further improvements of the accuracy are likely to be possible, by better optimization of the basis set, by simultaneously fine-tuning the representation of environments by $\lambda$-SOAP kernels and the representation of the density in terms of projections on a local basis set, by including self-consistent charge equilibration schemes~\cite{ghas+15prb} and also by using inexpensive semi-empirical methods to provide a baseline for the electron density prediction. 
In fact, this work can be seen as a first, successful attempt to apply machine learning in a transferable way to molecular properties that cannot be simply decomposed as the sum of atom-centered values, but exhibit a richer, more complex geometric structure. 
The Hamiltonian, the density matrix, vector fields and density response functions are other examples that will require careful consideration of both the representation of the input structure, and of the property one wants to predict, and that can benefit from the framework we have introduced in the present work.

\section*{Acknowledgments}

The authors are grateful to G. Cs\'anyi for insightful discussion.
M.C and D.M.W. were supported by the European Research Council under the European Union's Horizon 2020 research and innovation programme (grant agreement no. 677013-HBMAP), and benefited from generous allocation of computer time by CSCS, under project id s843.  A.G. acknowledges funding by the MPG-EPFL Center for Molecular Nanoscience and Technology. C.C., B.A.R.M. and A.F. acknowledge the National Centre of Competence in Research (NCCR) Materials Revolution: Computational Design and Discovery of Novel Materials (MARVEL) of the Swiss National Science Foundation (SNSF) for financial support and the EPFL for computing time.

\end{document}